\documentclass[showpacs,amsmath,amssymb,aps,pra,
10pt,reprint,superscriptaddress,showkeys]{revtex4-1}
\usepackage{graphicx}
\usepackage{bm}
\usepackage[breaklinks=true,colorlinks=true,linkcolor=blue,urlcolor=blue,citecolor=blue]{hyperref}
\usepackage{graphics}
\usepackage{dcolumn}
\usepackage{amsmath,amssymb}
\usepackage{mathdots}
\usepackage{natbib}
\usepackage{soul,color}
\usepackage{epstopdf}
\usepackage[note-name]{notes2bib}

\begin{document}
\title{\Large Ultra-fast vortex motion in dirty Nb-C superconductor\\ with a close-to-perfect edge barrier}

\author{O. V.~Dobrovolskiy}
    \email{oleksandr.dobrovolskiy@univie.ac.at}
    \affiliation{Faculty of Physics, University of Vienna, 1090 Vienna, Austria}
    \affiliation{Physics Department, V. Karazin National University, 61077 Kharkiv, Ukraine}
\author{D.~Yu.~Vodolazov}
    \affiliation{Institute for Physics of Microstructures, Russian Academy of Sciences, Nizhny Novgorod, Russia}
    \affiliation{Department of Physics and IT, Moscow Pedagogical State University, Moscow, Russia}
\author{F. Porrati}
\author{R. Sachser}
    \affiliation{Physikalisches Institut, Goethe University, 60438 Frankfurt am Main, Germany}
\author{V. M. Bevz}
    \affiliation{Physics Department, V. Karazin National University, 61077 Kharkiv, Ukraine}
\author{M. Yu. Mikhailov}
    \affiliation{B. Verkin Institute for Low Temperature Physics and Engineering of the National Academy of Sciences of Ukraine, Kharkiv, Ukraine}
\author{A. V.~Chumak}
    \affiliation{Faculty of Physics, University of Vienna, 1090 Vienna, Austria}
\author{M. Huth}
    \affiliation{Physikalisches Institut, Goethe University, 60438 Frankfurt am Main, Germany}
\date{\today}

\begin{abstract}
The ultra-fast dynamics of superconducting vortices harbors rich physics generic to nonequilibrium collective systems. The phenomenon of flux-flow instability (FFI), however, prevents its exploration and sets practical limits for the use of vortices in various applications. To suppress the FFI, a superconductor should exhibit a rarely achieved combination of properties: weak volume pinning, close-to-depairing critical current, and fast heat removal from heated electrons. Here, we demonstrate experimentally ultra-fast vortex motion at velocities of 10--15\,km/s in a directly written Nb-C superconductor in which a close-to-perfect edge barrier orders the vortex motion at large current values. The spatial evolution of the FFI is described using the edge-controlled FFI model, implying a chain of FFI nucleation points along the sample edge and their development into self-organized Josephson-like junctions (vortex rivers). In addition, our results offer insights into the applicability of widely used FFI models and suggest Nb-C to be a good candidate material for fast single-photon detectors.
\end{abstract}

\maketitle

\section{Introduction}

The dynamics of vortices at large transport currents is of major importance for the comprehension of vortex matter under far-from-equilibrium conditions and it sets practical limits for the use of superconductors in various applications \cite{Gur08prb,Che14apl,Gri15prb,Zel15nsr,Lar17pra,Ves16nac,Emb17nac,Mad18sca,Sha19met,Rou19nal}.
The physics of current-driven vortex matter is getting especially interesting when the vortex velocity exceeds the velocity $v\approx3$\,km/s of other possible excitations in the system, allowing for the Cherenkov-like generation of sound \cite{Ivl99prb,Bul05prb} and spin \cite{Bes13prb,Bes14prb} waves by moving fluxons, which opens up novel routes to excite waves in magnon spintronics \cite{Chu15nph,Jeo18nam,Boz20ltp,Moh20arx}. Furthermore, there is currently great interest in the interplay of Meissner currents and magnetic flux quanta with spin waves in the rapidly developing domain of magnon fluxonics \cite{Gol18afm,Dob19nph}, in which high vortex velocities are required for tuning the Bloch-like band structure of spin waves scattered on the moving vortex lattice.

The maximal current a superconductor can carry without dissipation is determined by the pair-breaking (depairing) current $I_\mathrm{dep}$. However, a highly-resistive state in real systems is usually attained at much smaller currents due to the presence of regions in which superconductivity breaks down long before $I_\mathrm{dep}$ is reached. Namely, in a vortex-free state, the earlier breakdown of superconductivity is due to spatial variations of the order parameter caused by structural imperfectnesses and the sample geometry \cite{Hen12prb,Cle11prb}. In the vortex state, fast-moving vortices are known to lead to a quench of the low-dissipative state at $I^\ast \ll I_\mathrm{dep}$ as a consequence of the flux-flow instability (FFI) associated with the escape of quasiparticles (normal electrons) from the vortex cores \cite{Lar76etp,Lar86inb}. Accordingly, to achieve $I_\mathrm{c} \lesssim I_\mathrm{dep}$ and high vortex velocities $v\gtrsim 5$\,km/s, a high structural homogeneity and fast cooling of quasiparticles (governed by the quasiparticles' energy relaxation time $\tau_\epsilon$ and the escape time of nonequilibrium phonons to the substrate $\tau_\mathrm{esc}$) are both required. However, while short $\tau_\epsilon$ is inherent to disordered superconducting systems \cite{Ili00apl,Zha18nsr}, few of them have $I_\mathrm{c} \lesssim I_\mathrm{dep}$ in conjunction with weak volume pinning needed to maintain long-range order in the fast-moving vortex lattice. Variation in the local pinning forces induced by uncorrelated disorder (volume pinning) leads to a broader distribution of $v$ and thereby prevents the exploration of vortex matter at high velocities \cite{Sil12njp,Shk17prb,Dob17sst,Bez19prb}.
Near the superconducting transition temperature, the FFI is described by the Larkin-Ovchinnikov (LO) mechanism associated with a shrinkage of the vortex cores and leading to an avalanche-like increase of the vortex velocity in consequence of a drastic reduction of the vortex viscosity coefficient \cite{Lar76etp,Lar86inb}. The LO model was developed for vortices moving with the same velocity, while a broad distribution of $v$ is typical for spatially inhomogeneous systems \cite{Sil12njp,Bez19prb}. The broadening of the $v$ distribution implies a sizable separation between the measured average velocity $\langle v \rangle$ and the maximal attainable velocity $v_\mathrm{max}$ as the avalanche-like onset of the FFI occurs as soon as the critical velocity $v^\ast$ is achieved even by a small number of faster-moving vortices \cite{Sil12njp,Bez19prb}.

Recently, two approaches were used to demonstrate ultra-fast vortex motion at $v \gtrsim 5$\,km/s. In the first case, a clean Pb bridge with both, an edge barrier for vortex entry and a high demagnetization factor (so-called geometrical barrier) was studied \cite{Emb17nac}. In the used geometry there was a strongly nonuniform current distribution both across and along the bridge due to a small Pearl length $2\lambda^2/d \ll w$, where $d$ and $w$ are the film thickness and width, respectively. A weak pinning in and a short electron-phonon relaxation time $\tau_\mathrm{ep}$ in Pb \cite{Wat81ltp} allowed one to diminish nonequilibrium effects and achieve the regime with ultra-fast Abrikosov-Josephson vortices \cite{Emb17nac}. In the second case, an array of ferromagnetic Co nanostripes on top of a superconducting Nb film led to a dynamic ordering of flux quanta guided by the nanostripes and allowed to achieve a narrow distribution of their velocities \cite{Dob19pra}. In both of these approaches, specially-designed, locally non-uniform structures were used. At the same time, a close-to-ideal uniform system where the fast heat removal from electrons rather than the finite width of the $v$ distribution becomes the limiting factor for ultra-fast vortex dynamics was never investigated experimentally. Theoretically, however, it was recently predicted that dirty superconductors with weak volume pinning and strong edge barrier for vortex entry should also allow for ultra-fast vortex dynamics \cite{Vod19sst}. Extremely dirty superconductors are known to have a short electron-electron inelastic scattering time $\tau_\mathrm{ee}$ which leads to a decrease of $\tau_\mathrm{ep}$ \cite{Vod17pra}. This diminishes nonequlibrium effects and may lead to an increase of the critical velocity of vortices. One of the most important requirements for the observation of an edge-controlled FFI is a spatially homogenous edge in conjunction with a weak pinning in the superconductor's volume \cite{Vod19sst}. The presence of a strong edge barrier in such superconductors leads to a current gradient near the edge where vortices enter the superconductor and where FFI is actually nucleating.

Here, we demonstrate experimentally the phenomenon of \emph{edge-barrier-controlled flux-flow instability} in direct-write superconductors with a close-to-perfect edge barrier and deduce vortex velocities up to $15$\,km/s from their current-voltage curves ($I$-$V$). The investigated system is the recently synthesized Nb-C superconductor fabricated by focused ion beam induced deposition (FIBID) \cite{Por19acs}, with a very high resistivity $\rho =572\,\mu \Omega$cm. This implies a large effect of the inelastic electron-electron scattering with the characteristic times $\tau_\mathrm{ee} \lesssim \tau_\mathrm{ep}$ which speeds up the relaxation of disequilibrium. The Nb-C microstrips have a rather low depinning current and their critical current is controlled by the edge barrier for vortex entry. In contrast to Ref. \cite{Emb17nac}, in our system $\lambda^2/d \gg w$, which means a negligible demagnetization factor (no geometrical barrier) and a uniform current distribution across the strip at zero magnetic field. The spatial evolution of the FFI is described in terms of the edge-barrier-controlled FFI model recently developed by one of the authors \cite{Vod19sst}, implying a chain of FFI nucleation points along the sample edge and their development into self-organized Josephson-like junctions (vortex rivers) evolving to normal domains which expand along the entire sample. In addition, our results offer insights into the applicability of widely used FFI models and render Nb-C to be a good candidate material for fast single-photon detectors.

\section{Results}
\begin{figure}[b!]
    \centering
    \includegraphics[width=0.79\linewidth]{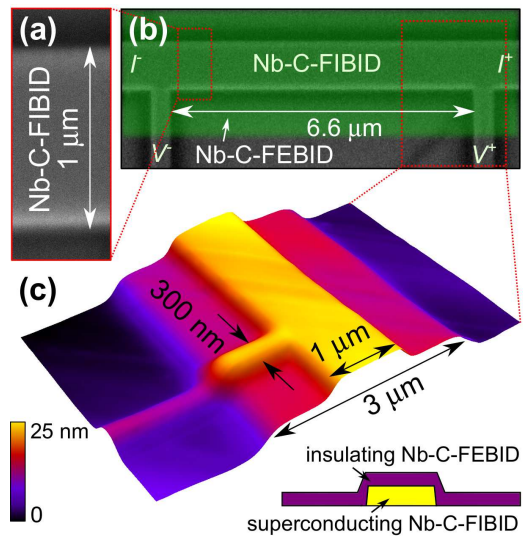}
    \caption{\textbf{Experimental geometry}.
    Scanning electron microscopy images of the superconducting Nb-C-FIBID microstrip before (a) and after (b) covering it with an insulating Nb-C-FEBID layer shown by the green false-color.
    The current and voltage leads are indicated with $I^+$, $I^-$, $V^+$, and $V^-$. (c) Atomic force microscopy image of a part of the fabricated structure.}
    \vspace{3mm}
    \label{f1}
\end{figure}
\textbf{System under investigations}. We study the vortex dynamics in a direct-write Nb-C superconducting microstrip fabricated by focused ion beam induced deposition (FIBID) \cite{Por19acs}. The microstrip is characterized by a transition temperature of $T_\mathrm{c} = 5.6$\,K and close-to-depairing values of the zero-field critical current $I_\mathrm{c}\approx0.7-0.74 I_\mathrm{dep}$ above $0.5T_\mathrm{c}$. The dimensions of the microstrip are: thickness $d = 15$\,nm, width $w = 1\,\mu$m, and length $l = 6.6\,\mu$m, see Fig.\,\ref{f1} for the geometry. The perpendicular-to-film-plane magnetic field with induction $\mathbf{B} = \mu_0\mathbf{H}$ populates the microstrip with a lattice of Abrikosov vortices. The applied dc current exerts a Lorentz force on the vortices that causes their motion with velocity $v$ across the microstrip. The associated voltage drop $V$ along the microstrip is recorded a function of the applied current $I$ in the current-biased mode. The microstrip is capped with an insulating Nb-C layer fabricated by focused electron beam induced deposition (FEBID) \cite{Por19acs,Hut18mee}. Further details on the sample fabrication and its structural properties are given in the Methods section.

\begin{figure}[t!]
    \centering
    \includegraphics[width=0.95\linewidth]{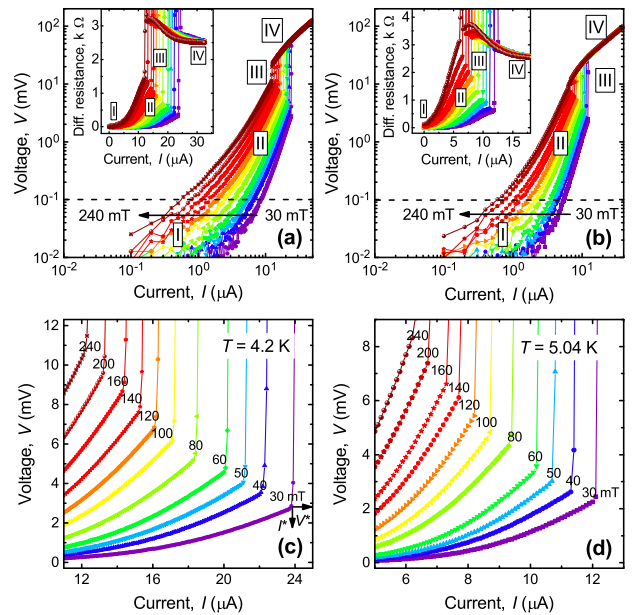}
    \caption{\textbf{Current-voltage curves of the Nb-C-FIBID microstrip}.
    $I$-$V$ curves of the microstrip in a series of magnetic fields at temperatures as indicated in panels (c) and (d). The different resistive regimes are indicated: pinned vortices (I), nonlinear conductivity in the flux flow regime (II), flux-flow instability (III), and the normal state (IV). The instability jumps are enlarged in panels (c) and (d). The arrows in (c) illustrate the definition of the instability current $I^\ast$ related to the instability voltage $V^\ast$.}
    \vspace{3mm}
    \label{f3}
\end{figure}

\begin{figure}[t!]
    \centering
    \includegraphics[width=0.95\linewidth]{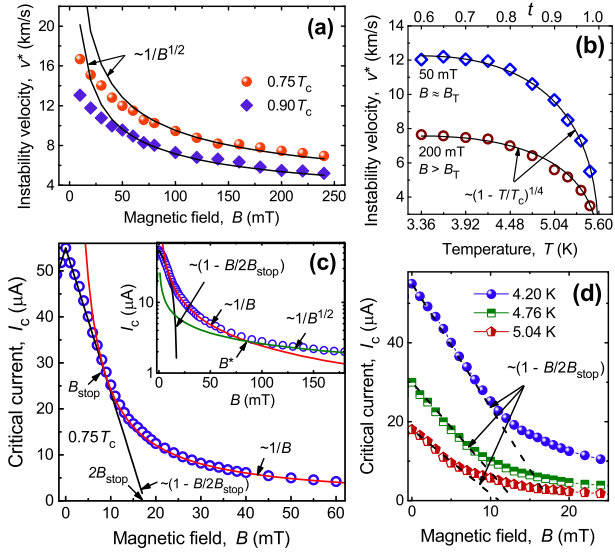}
    \caption{\textbf{Instability velocity and critical current in the microstrip}.
    (a) Instability velocity $v^\ast$ as a function of the magnetic field. Symbols: Experiment. Solid lines: fits to Eq. (\ref{eHub}).
    (b) Temperature dependence of the instability velocity at $50$\,mT and $200$\,mT. Symbols: Experiment. Solid lines: fits to Eq. (\ref{evInst}).
    (c) Crossover from the linear dependence $I_\mathrm{c}(B)$ at $B<B_\mathrm{stop}$ to $I_\mathrm{c}(B)  \sim1/B$ for $B_\mathrm{stop}< B < B^\ast$
    and $I_\mathrm{c}(B)\sim1/\sqrt{B}$ for $B>B^\ast$ at $4.20$\,K. The inset shows the same data in $\log (I_\mathrm{c})$ versus $B$ representation.
    (d) Dependence of the critical current $I_\mathrm{c}$ of the microstrip on the magnetic field at three different temperatures, as indicated.
    }
    \label{fVelocity}
\end{figure}

\textbf{Current-voltage characteristics.} Figure \ref{f3} displays the $I$-$V$ curves measured at $4.2\,$K ($0.75T_\mathrm{c}$) and $5.04$\,K ($0.9T_\mathrm{c}$) for a series of magnetic fields between $30$\,mT and $240$\,mT. With increase of the current, a series of different resistive regimes can be identified, as indicated in the $I$-$V$ curves: the pinned regime (I), the nonlinear flux-flow regime (II), and the FFI (III) causing abrupt onsets of the normal state (IV). Of especial interest for the following is the regime of high vortex velocities just before the FFI (III) with the $I$-$V$ sections enlarged in Fig.\,\ref{f3}(c) and (d). From the last points before the voltage jumps, referring to Fig.\,\ref{f3}(c) and (d), the vortex instability velocity $v^\ast$ is deduced by the relation $v^\ast = V^\ast/(B L)$. The resulting dependence $v^\ast(B)$ is presented in  Fig.\, \ref{fVelocity}(a). Remarkably, $v^\ast$ is between $5$\,km/s and $10$\,km/s at larger fields $B \gtrsim 100$\,mT and it is between $10$\,km/s and $15$\,km/s at $B < 100$\,mT. The temperature dependence $v^\ast(T)$ is presented in Fig.\,\ref{fVelocity}(d) for two magnetic field values. The field $50$\,mT is exemplary for a relatively sparse vortex lattice (vortex lattice spacing $a\approx220$\,nm) while $a \approx 110$\,nm at $200$\,mT for the assumed triangular vortex lattice with $a=\sqrt{2\Phi_0/\sqrt{3}H}$, where $\Phi_0$ is the magnetic flux quantum. At both fields, the experimental data nicely fit the law $v^\ast \sim (1 - t)^{1/4}$, where $t = T/T_\mathrm{c},$ with $v^\ast(0.6T_\mathrm{c}, 50\,\mathrm{mT}) = 12$\,km/s and $v^\ast(0.6T_\mathrm{c}, 200\,\mathrm{mT}) = 7.7$\,km/s, while a deviation of $v^\ast(T)$ from the $B^{-1/2}$ dependence is observed at $B \lesssim 50$\,mT in Fig.\,\ref{fVelocity}(a). The decreasing dependence of $v^\ast(B)$ below about $10$\,mT due to the decreasing vortex density (the so-called low-field crossover in the $v^\ast(B)$ dependence \cite{Gri10prb}) is beyond our consideration, as we are especially interested in the regime of very high vortex velocities.

\textbf{Influence of the edge barrier on the vortex dynamics.} The magnetic field dependence of the critical current at $4.20$\,K is presented in Fig.\,\ref{fVelocity}(c). At smaller fields, $I_\mathrm{c}(B)$ decreases linearly with $B$, while at larger fields the decrease of $I_\mathrm{c}$ becomes nonlinear and slower. This behavior can be explained by the presence of some threshold field $B_\mathrm{stop}$, which demarcates the Meissner (vortex free) and the mixed states of a superconducting stripe \cite{Mak98pss}. Namely, the dependence $I_\mathrm{c}(B)$ in the Meissner state ($B < B_\mathrm{stop}$) is linear and it is described by the expression $I_\mathrm{c}(B) = I_\mathrm{c}(0\,\mathrm{T})(1- B/2B_\mathrm{stop})$, where $B_\mathrm{stop}$ in the Ginzburg-Landau model \cite{Mak01epl} is given by $B_\mathrm{stop} = B_\mathrm{s}/2 = \Phi_0/(2\sqrt3\pi\xi(T) w)$. Here, $B_\mathrm{s}$ is the field value at which the surface barrier for vortex entry is suppressed at $I = 0$, $\xi$ is the superconducting coherence length, and $w$ is the microstrip width. The definition of $B_\mathrm{stop}$ following from $I_\mathrm{c}(2B_\mathrm{stop}) =0$ is illustrated in Fig.\,\ref{fVelocity}(c).

For $10\,\mathrm{mT}\lesssim B\lesssim 100$\,mT, the dependence of the critical current is described well by the dependence $I_\mathrm{c}(B) = I_\mathrm{c}(0\,\mathrm{T})B_\mathrm{stop}/2B$, which is the fingerprint of the edge mechanism of vortex pinning. At larger fields, $B\gtrsim 100$\,mT, a further crossover at $B^\ast$ to a slower decrease of $I_\mathrm{c}(B)$ as $1/\sqrt{B}$ is observed. This dependence can be explained by the increasing role of the intrinsic pinning at higher vortex densities at larger magnetic fields. The assumed origin of the intrinsic pinning is the order parameter suppression at the grain boundaries of individual crystallites in the Nb-C-FIBID microstrip.

The $I_\mathrm{c}$ values have been checked to be almost independent of the microstrip width for $w = 250-1000$\,nm at fields $B_\mathrm{stop} \lesssim B  \lesssim B^\ast$, that further corroborates the decisive role of the edge barrier for vortex pinning. Indeed, from the inset in Fig. \ref{fVelocity}(c) follows that at $B>150$\,mT the difference between the experimentally measured critical current $I_\mathrm{c}$ and the critical current calculated within the framework of the edge barrier model yields critical current densities for the volume pinning smaller than $6$\,kA/cm$^2$, being a factor of $50$ smaller than the critical current density at $B=0$. With a decrease of the width of the superconducting strip, the experimental data better follow the $1/B$ law up to larger fields (not shown) that can be understood as a consequence of the decreasing volume pinning contribution with respect to the edge barrier pinning. Thus, at vortex velocities $v^\ast\backsimeq10$\,km/s, the pronounced edge barrier for vortex entry is expected to have a strong impact on the FFI \cite{Vod19sst}, as will be discussed next.

\section{Discussion}
\textbf{Applicability of the modified Larkin-Ovchin\-nikov theoretical models.} Prior to a discussion of the obtained results it is worth to summarize the most important experimental findings, namely: (i) high vortex velocities $v>5$\,km/s in magnetic fields below $240$\,mT, and (ii) the $I_\mathrm{c}(B) \thicksim1/B$ dependence at $B<100$\,mT (where the highest vortex velocities are observed) pointing to the dominating edge mechanism of vortex pinning. Accordingly, further insights into the spatial evolution of the FFI can be provided by an edge-barrier-controlled model \cite{Vod19sst} of FFI which was recently developed for strongly disordered superconductors with weak volume pinning and strong edge barrier for vortex entry. However, we first compare the experimental results with the well-known and widely used Larkin-Ovchinnikov (LO) model \cite{Lar75etp,Lar86inb} with the modifications introduced by Bezuglyj and Shklovskij (BS) \cite{Bez92pcs} and Doettinger \emph{et al} \cite{Doe95pcs}. Although edge barrier effects are not considered in these models \cite{Lar75etp,Lar86inb,Bez92pcs,Doe95pcs}, it is still interesting to check what quasiparticle energy relaxation time $\tau_{\epsilon}$ values, related to the instability velocity, can be deduced from fitting of the experimental data to these models.

Within the framework of the LO theory \cite{Lar75etp,Lar86inb}, the microscopic mechanism of FFI is the following. When the electric field induced by vortex motion raises the quasiparticle energy above the potential barrier associated with the order parameter around the vortex core, quasiparticles leave it and the core shrinks. The shrinkage of the vortex cores leads to a reduction of the viscous drag coefficient and a further avalanche-like acceleration of the vortex, eventually quenching the low-resistive state. The original LO theory was developed in the dirty limit near $T_\mathrm{c}$ and in neglect of heating of the superconductor. To account for quasiparticle heating due to the finite heat-removal rate of the power dissipated in the sample, the LO theory was extended by BS \cite{Bez92pcs}. In the BS generalization, the latter effect was considered in the framework of the kinetic equation LO approach, which assumes a non-thermal (non-Fermi-Dirac) electron distribution function, while Joule heating was taken into account using the thermal distribution function and the electron temperature $T_\mathrm{e}$ was determined from the heat conductance equation. In contrast to the $B$-independent instability velocity $v^\ast$ in the LO model, a $v^\ast(B)$ variation is expected in the BS model \cite{Bez92pcs} and takes the form
\begin{equation}
\label{evInst}
    v^\ast \propto h(1 - t)^{1/4} B^{-1/2},
\end{equation}
where $h$ is the heat removal coefficient. While the magnetic field dependence $v^\ast(B)$ nicely fits Eq. (\ref{evInst}) at $B\gtrsim 50$\,mT, a notable deviation of $v^\ast(B)$ towards smaller values is observed in Fig. \ref{fVelocity}(a) at $B\lesssim 50$\,mT. This deviation will be commented in the next subsection. In all, the complete set of the instability parameters deduced from Fig. \ref{f3} nicely fits the BS scaling law (see Appendix). However, if one associates $\tau_{\epsilon}$ with the electron-phonon scattering time $\tau_\mathrm{ep}$ in the LO model, the deduced $\tau_{\epsilon}$ is at least one order of magnitude smaller than one could expect from $\tau_{\epsilon}$ found in similar low-$T_\mathrm{c}$ highly disordered superconductors \cite{Bab04prb,Sid18prb,Sid19arx}.

In the LO model modified by Doettinger \emph{et al} \cite{Doe95pcs} \cite{Doe95pcs,Leo11prb}, the quasiparticle energy relaxation time can be found from the following equation
\begin{equation}
\label{eHub}
    v^\ast = \left[\frac{(1-t)^{1/2}D[14\zeta(3)]^{1/2}}{\pi \tau_\epsilon}\right]^{1/2}
    \left(1 + \frac{a}{\sqrt{D \tau_\epsilon}}\right).
\end{equation}
In Eq. (\ref{eHub}), the term $a/\sqrt{D \tau_\epsilon}$, where $a$ is the intervortex distance, has been added to incorporate the necessary condition of spatial homogeneity of the nonequilibrium quasiparticle distribution between vortices at relatively small magnetic fields. The calculation results by Eq. (\ref{eHub}) are shown by solid lines in Fig.\,\ref{fVelocity}(a) where the energy relaxation time has been varied as the only fitting parameter. The best fits are achieved with $\tau_\epsilon=16$\,ps which could be considered as a more accurate estimate for the energy relaxation time in the Nb-C-FIBID superconductor. We note that with this $\tau_\epsilon$ estimate, the quasiparticle diffusion length $l_\epsilon = \sqrt{D \tau_\epsilon} \approx 28$\,nm is much smaller than the intervortex distance $a$ at all used magnetic fields and, importantly, $l_\epsilon\lesssim 2\xi(T)$ with $2\xi(0.75T_\mathrm{c})\approx 25$\,nm and $2\xi(0.9T_\mathrm{c})\approx 38$\,nm.

\textbf{Description of the experimental results by the edge-barrier-controlled instability model.}
\begin{figure*}[t!]
    \centering
    \includegraphics[width=0.85\linewidth]{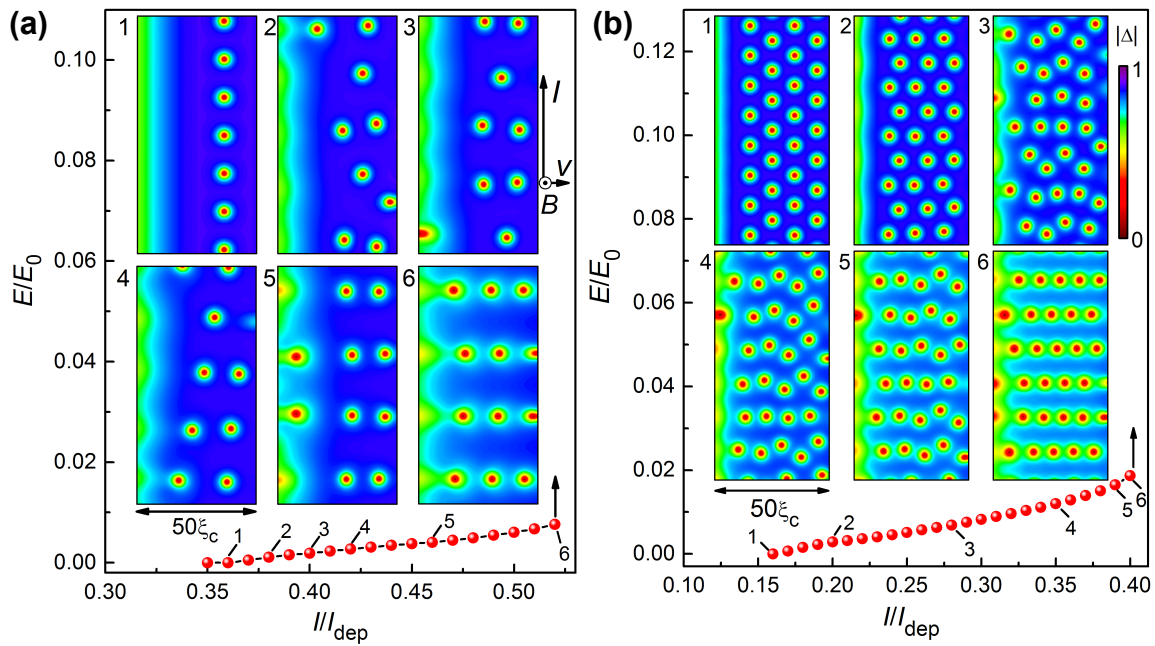}
    \caption{\textbf{Time-dependent Ginzburg-Landau simulations.} Calculated $I$-$V$ curves of a superconducting microstrip with width $w = 50\xi_\mathrm{c}$ at $T = 0.8T_\mathrm{c}$ for $B = 0.02\,B_\mathrm{0}$ (a) and $B = 0.05\,B_\mathrm{0}$ (b).
    The insets show snapshots of the superconducting order parameter $|\Delta|$ at different current values, as indicated. For the studied system, the parameter $B_0 = \Phi_0/(2\pi\xi_\mathrm{c}^2) \simeq 4.9$\,T, where $\xi_\mathrm{c}= \sqrt{1.76}\xi(0) = 8.2$\,nm. The electric field is measured in units of $E_0 = k_\mathrm{B}T_\mathrm{c}/(2e\xi_c)$ and the current in units of $I_\mathrm{dep}$.}
    \vspace{6mm}
    \label{fTDGL}
\end{figure*}
The edge-barrier-controlled FFI scenario \cite{Vod19sst} is different from the FFI scenario of LO and BS. Indeed, LO and BS considered a moving periodic vortex lattice in an infinite superconductor in the Wigner-Seitz approximation and hence could not take into account the collective effects related to the transformation of the vortex lattice and edge barrier effects. In contrast, in the edge-barrier-controlled FFI model \cite{Vod19sst} these effects are taken into account, as well as the local Joule heating and cooling (due to the time variation of the magnitude of the superconducting order parameter $|\Delta|$) depending on the vortex position. The edge-barrier-controlled FFI model allows to study a ``local'' instability and collective effects in the vortex dynamics relying upon the solution of a heat conductance equation for the electrons and a modified time-dependent Ginzburg-Landau equation for $\Delta(r,t)$. In this model, it was shown that, in the low-resistive state, there is a temperature gradient across the width of the microstrip with maximal local temperature near the edge where vortices enter the sample \cite{Vod19sst}. The higher temperature at the edge is caused by the larger current density in the near-edge area due to the presence of the edge barrier for vortex entry and, hence, the locally larger Joule dissipation. With increase of the current, there is a series of transformations of the moving vortex lattice. In Fig. \ref{fTDGL} we show examples of the calculated $I$-$V$ curves and snapshots of $|\Delta|(r)$ for the parameters of the superconductor as in Ref. \cite{Vod19sst}. Similar transformations connected with reorientations of the moving vortex lattice in the insets 1-2 in Fig.\,\ref{fTDGL}(b) were experimentally observed in Ref. \cite{Oku12prb} and theoretically analyzed in Ref. \cite{Vod07prb}.

At currents just below $I^\ast$, localized areas with strongly suppressed superconductivity and closely spaced vortices appear near the hottest edge (left edge in the insets in Fig.\,\ref{fTDGL}). Upon reaching $I^\ast$, these areas begin to grow in the direction of the opposite edge and form a highly resistive Josephson SNS-like link (vortex river) along which vortices move \cite{Vod19sst,Sil10prl,Gri15prb,Emb17nac}. These vortices are of the Abrikosov-Josephson type, as they are moving in areas with suppressed order parameter. Due to the increasing dissipation, vortex rivers evolve into normal domains which than expand along the microstrip. In consequence of this, a jump to the highly resistive state occurs at $I^\ast$. In all, the simulation results demonstrate that transformation of the moving vortex array is a collective phenomenon, which involves correlated changes in the motion of many vortices with increase of the current and, at $I^\ast$, results in the appearance of Josephson-like SNS links known as vortex rivers \cite{Sil10prl,Gri15prb,Emb17nac}.
\begin{figure*}[t!]
    \centering
    \includegraphics[width=0.8\linewidth]{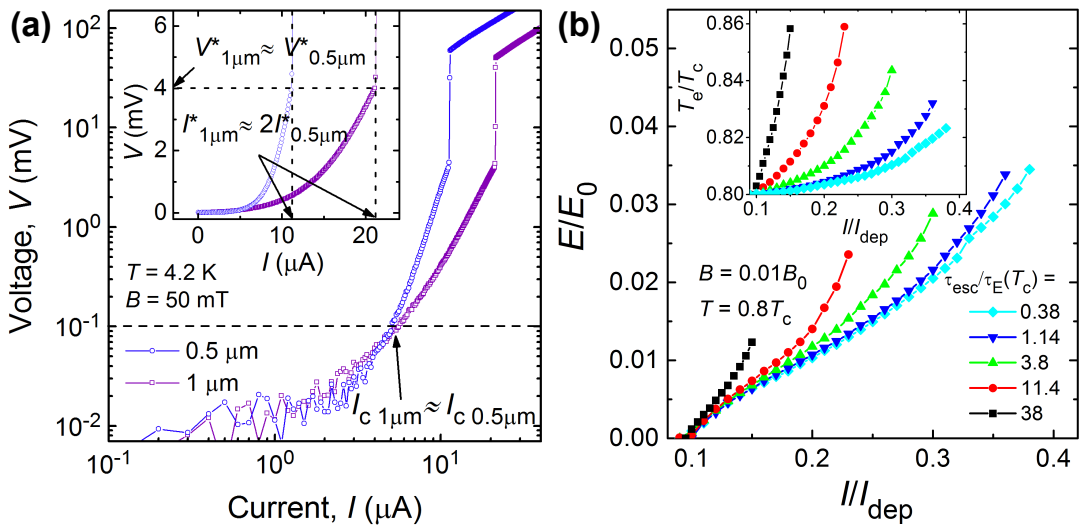}
    \caption{\textbf{Current-voltage curves of the microstrips.}
    (a) Experimental $I$-$V$ curves of the two Nb-C-FIBID microstrips with the widths $w = 1\,\mu$m and $500$\,nm at $T = 4.2$\,K and $B = 50$\,mT in the double log representation. Inset: the same data in the linear scale.
    (b) Calculated $I$-$V$ curves of a superconducting microstrip with the width $w = 50\xi_\mathrm{c}$ at $T = 0.8T_\mathrm{c}$, $B = 0.01\,B_\mathrm{0}$ and for different $\tau_\mathrm{esc}$ values, as indicated. Inset: time-averaged electronic temperature $T_\mathrm{e}$ in the center of the microstrip as a function of the normalized current. Parameters are as in Ref. \cite{Vod19sst}: $C_\mathrm{e}(T_\mathrm{c})/C_\mathrm{p}(T_\mathrm{c})=0.57$, $\tau_\mathrm{E} =12.5$\,ps, $\tau_\mathrm{E}(0.8T_\mathrm{c})\simeq 2\tau_\mathrm{E}(T_\mathrm{c})$.}
    \vspace{5mm}
    \label{fCVCw}
\end{figure*}

An interesting feature of the snapshots of the order parameter in the insets of Fig. \ref{fTDGL}(a) is that the number of vortices in the microstrip is by up to $40\%$ smaller than it would follow from the simple estimate $n \Phi_0 = B S$, where $n$ is the number of vortices and $S$ is the sample area. This difference is connected with the combined effect of the current and the edge barrier that decreases the number of vortices below its equilibrium value. However, at fields $B \gg B_\mathrm{stop}$ this discrepancy becomes smaller and at $B=0.05 B_0 \sim 2B_\mathrm{stop}$ it amounts to about $25\%$. The smaller number of vortices in the sample following from the simulations suggests that the actual $v^\ast$ values may be even higher if one substituted the correct number of vortices at the instability point.

An important check which should be done to further justify the use of the edge-controlled FFI model
concerns the dependence of the instability current on the microstrip width $w$. Namely, in the edge-controlled FFI model \cite{Vod19sst} the current $I^{\ast}$ increases linearly with the width of the strip while $V^{\ast}$ does not depend on $w$ as it does in the LO and BS models. This result holds at $B \gg B_\mathrm{stop}$ when $a$ is much smaller than the microstrip width $w$ and $a$ becomes smaller than the width of the vortex-free region near the edge of the microstrip. This means that despite the nucleation of FFI points occurs near the edge where the local temperature and the current densities are maximal, far from the edge where the current density is uniform, the vortices should move at relatively high velocities. Otherwise the FFI will not develop across the whole microstrip and one has only origins of the vortex rivers, as it can be seen from Fig.\,5 in \cite{Vod19sst} at $I\lesssim I^{\ast}$. The linear scaling of $I^\ast(w)$ with the microstrip width $w$ is corroborated by the experimental observation in Fig. \ref{fCVCw}(a), where the $I$-$V$ curves for two microstrips with the widths $w = 1\,\mu$m and $500$\,nm are shown at $T = 4.2$\,K and $B = 50$\,mT.

In the edge-barrier-controlled FFI model \cite{Vod19sst}, the energy relaxation time depends not only on the electron-phonon relaxation time $\tau_\mathrm{ep}$ (as in the LO model) but also on the escape time of nonequilibrium phonons to the substrate $\tau_\mathrm{esc}$ and the ratio of the electron and phonon heat capacities, $C_\mathrm{e}$ and $C_\mathrm{p}$, respectively. At $T\backsimeq T_\mathrm{c}$ and for a small deviation from equilibrium one has
\begin{equation}
    \tau_{\epsilon} \simeq \tau_\mathrm{E}+\tau_\mathrm{esc}(1+C_\mathrm{e}(T_\mathrm{c})/C_\mathrm{p}(T_\mathrm{c})).
    \label{tau_e}
\end{equation}
where $\tau_\mathrm{E} \simeq \tau_\mathrm{ep}/4.5$ is the electron-phonon relaxation time renormalized due to fast electron-electron inelastic scattering. Here, $\tau_\mathrm{ep}$ is the electron-phonon relaxation time used in the LO model. Following the arguments of Ref. \cite{Doe95pcs} one can claim that the instability occurs at the velocity $v^\ast \sim a/\tau_\mathrm{\epsilon}$ when the intervortex distance $a\lesssim\sqrt{D\tau_\mathrm{\epsilon}}$. This condition leads to a dependence of $v^\ast(B)$, which was revealed in numerical calculations \cite{Vod19sst}. One important difference between the modified LO model \cite{Doe95pcs} and the edge-controlled FFI model is that in the latter \cite{Vod19sst}, $a \sim B^{-1/2}$ only at relatively large magnetic fields, when the intervortex distance at $I\sim I_\mathrm{c}$ and $I\sim I^{\ast}$ is almost the same despite the change in the structure of the moving vortex lattice. At relatively small magnetic fields, $a$ in the vortex rows is smaller than $(2\Phi_0/B\sqrt{3})^{1/2}$ at $I\sim I^{\ast}$ and, thus, the number of vortices is smaller than follows from the simple estimate $n \Phi_0 = B S$, see Fig.\,\ref{fTDGL}(a). Altogether, this leads to a weaker experimental dependence $v^{\ast}(B)$ than follows from the ``global'' instability model with $v^{\ast} \sim B^{-1/2}$ \cite{Doe95pcs}. Qualitatively, it is this behavior which is observed in the experiment, see Fig. \ref{fVelocity}(a).
\begin{table*}[t!]
    \centering
    \begin{tabular}{|l|c|c|c|c|c|c|}
    \hline
    Material                             &  MoSi \cite{Kor20pra} & NbRe \cite{Cap17apl} & NbN \cite{Kor18pra}& NbN \cite{Sem09prb}& NbC \cite{Kor18pra}& Nb-C-FIBID \\
    \hline
    $d$, nm                              & 3.3                   & 15                   & 5.8                & 14.4               & 23.3                 & 15\\
    \hline
    $T_\mathrm{c}$,\,K                    & 3.85                  & 6.77                 & 8.35               & 15.25              & 11.2                & 5.6\\
    \hline
    $\rho_\mathrm{n}$,\,$\mu\Omega$cm     & 175                   & 145                  & 400                & 281                 & 25                & 572\\
    \hline
    $D$,\,cm$^2$/s                       & 0.47                  & 0.56                  & 0.31               & 0.6                & 4.45              & 0.49\\
    \hline
    $\lambda(0)$,\,nm                    & 708                   & 483                   & $450^\ast$         & $290^\ast$          & $156^\ast$           & 800\\
    \hline
    $\xi(0)$,\,nm                        & 8.7                   & 4.8                   & 5.4                & 5.4                & $-$                 & 8\\
    \hline
    $I_\mathrm{c}/I_\mathrm{dep}$       & $\simeq0.7$           & $-$                 & $-$              & $0.65-0.9$         &  $-$         & $0.7-0.74$\\
    \hline
    Leads                               & tapered               & straight              & tapered           & straight            &  straight            & straight\\
    \hline
    \end{tabular}
    \caption{Comparison of Nb-C-FIBID with some superconducting materials used for single-photon detection.
    $d$: stripe thickness; $T_\mathrm{c}$: superconducting transition temperature; $\rho_\mathrm{n}$ resistivity just above $T_\mathrm{c}$;
    $\lambda(0)$ estimate for the penetration depth at zero temperature; $\xi(0)$: zero-temperature estimate for the coherence length.
    The asterisk $^\ast$ denotes an estimate which was made on the basis of the reported data.
    \vspace{3mm}
    \label{t1}}
\end{table*}

We finally note that in dirty superconductors with short electron-electron scattering time $\tau_\mathrm{ee} \ll \tau_\mathrm{ep}$, $\tau_{\epsilon}$ is smaller by a factor of $4.5$ than in superconductors with $\tau_\mathrm{ee} > \tau_\mathrm{ep}$. Therefore one can expect large $v^{\ast}$ values at given $\tau_\mathrm{ep}$ and other parameters. However, there is an additional term in Eq. (\ref{tau_e}), the escape time of nonequilibrium phonons to the substrate, $\tau_\mathrm{esc}$, which should also be small in comparison to $\tau_\mathrm{E}$ to achieve high vortex velocities. If this is not the case, than $v^{\ast}$ can be even smaller than in moderately dirty superconductors. For example, extremely large $\tau_{\epsilon}> 1$\,ns were deduced for NbN and Mo$_3$Si from FFI analysis within the framework of the LO model \cite{Sam95prl,Lin13prba}, pointing to a large contribution of $\tau_\mathrm{esc}$ to $\tau_{\epsilon}$ in those works.

In this way, the large $v^{\ast}$ values observed in our system should be attributed not only to $\tau_\mathrm{E}<\tau_\mathrm{ep}$ but to a small $\tau_\mathrm{esc}$ as well. Indeed, due to the insulating Nb-C-FEBID layer on top of the microstrip, there seems to be no phonon bottleneck which could exist due to an acoustic mismatch between a thin dirty superconductor and a substrate \cite{Sid18prb}. As an estimate, for our system we deduce $\tau_\mathrm{esc} \sim 4d/u \approx 24$\,ps, where $u \sim 2.5$\,km/s is the mean sound velocity. This value is larger than $\tau_{\varepsilon} \sim 16$\,ps deduced from the experimental data using the modified LO model. We have to stress that numerical coefficients in the LO model are strictly valid only rather close to $T_\mathrm{c}$ (when $\Delta(T) \ll k_\mathrm BT_\mathrm{c}$, i.e. at $T\gtrsim 0.9 T_\mathrm{c}$) and in the case when $\tau_\mathrm{ee}\gg
\tau_\mathrm{ep}$ and $\tau_\mathrm{esc}=0$. Therefore these coefficients may be different in our dirty system with $\tau_{\epsilon}\sim \tau_\mathrm{esc}$ and at temperatures further away from $T_\mathrm{c}$.

Finally, we would like to note that, unfortunately, there is no analytical relation between $v^{\ast}$ and $\tau_{\epsilon}$ in the edge-barrier-controlled FFI model \cite{Vod19sst}. Accordingly, a discussion of the relation between $v^{\ast}$ and $\tau_{\epsilon}$ has to remain on a qualitative level. From Eq. (\ref{tau_e}) it follows that a change of $\tau_\mathrm{E}$, $\tau_\mathrm{esc}$ and $C_\mathrm{e}/C_\mathrm{p}$, leads to a change of the relaxation time $\tau_{\epsilon}$. To illustrate this, in Fig.\,\ref{fCVCw}(b) we present a series of calculated $I$-$V$ curves at different $\tau_\mathrm{esc}$ values while the other parameters are kept fixed. Indeed, with increasing $\tau_\mathrm{esc}$ the critical velocity $v^{\ast} \sim E^{\ast}$ decreases, but it decreases slower than $\tau_{\epsilon}^{-1}$ or $\tau_{\epsilon}^{-1/2}$. Qualitatively, the same tendency is found if one increases the ratio $C_\mathrm{e}/C_\mathrm{p}$ for a given $\tau_\mathrm{esc}$ value. Specifically, with an increase of $\tau_\mathrm{esc}/\tau_\mathrm{E}$ by two orders of magnitude, $E/E_0$ decreases by only about a factor of three. In the inset of Fig.\,\ref{fCVCw}(b) one can also see that with increase of $\tau_\mathrm{esc}$, the time-averaged temperature in the center of the superconducting microstrip increases, that indicates an increased contribution of Joule dissipation to the FFI. The increased temperature also affects $v^{\ast}$ because of the temperature dependence $\tau_\mathrm{E} \sim 1/T^3$ and $C_\mathrm{e}/C_\mathrm{p} \sim 1/T^2$ in the used model \cite{Vod19sst}.

\textbf{Assessment of Nb-C-FIBID as a material for single-photon detectors}. We would like to outline an applications-related aspect of the superconducting properties of the studied Nb-C-FIBID microstrip. Namely, the small diffusivity $D\approx 0.49$\,cm$^2$/s and the low transition temperature $T_\mathrm{c} = 5.6$\,K suggest that Nb-C-FIBID may be a candidate material for SSPDs. We refer to Table \ref{t1} for a comparison with parameters of some typical SSPDs and to Ref. \cite{Vod17pra} for a further discussion. In this regard, it should be mentioned that for about a decade SSPDs were made of meandering nanostrips with widths in the range $50$ to $150$\,nm as it was empirically found that the use of wider strips leads either to the loss of the single-photon nature of the response or to a decrease of the detection efficiency \cite{Nat12sst}. This observation was in line with a ``geometric-hot-spot'' detection model, in which the width of the supercurrent-carrying strip should be comparable with the diameter of the normal region where the superconducting state is suppressed due to the absorption of the photon.

Recently, a ``photon-generated superconducting vortex model'' was proposed \cite{Zot12prb,Vod17pra}. It was revealed that the efficiency of the photon detection is not determined by the geometry, as long as the initial current density is uniform and close to the critical pair-breaking current $I_\mathrm{dep}$. It was emphasized that even several micron wide dirty superconducting stripes should be suitable to detect single near-infrared or optical photons if their critical current $I_\mathrm{c}\gtrsim 0.7I_\mathrm{dep}$ \cite{Vod17pra}. The only requirement for the width of the strip is that it should be smaller than the Pearl length $\Lambda = 2 \lambda^2/d$ that ensures the uniformity of the supercurrent across the superconductor width. Recently, this condition was satisfied in wide and short NbN \cite{Kor18pra} and MoSi \cite{Kor20pra} bridges, whose photon response was consistent with the vortex-assisted mechanism of initial dissipation \cite{Zot12prb}. In this way, given the superconducting properties of our samples, which drastically differ from much cleaner NbC films prepared by pulsed laser ablation in Ref. \cite{Kor16tas}, Nb-C-FIBID appears to be a good candidate for fast single-photon detection. A further enhancement of the critical current in Nb-C-FIBID can be expected for tapered current leads \cite{Kor18pra,Kor20pra} which should minimize the reduction of $I_\mathrm{c}$ in consequence of undesired current-crowding effects \cite{Cle11prb}, and additional advantages of easy on-chip \cite{Kah15nsr} or on-fiber \cite{Bac12apl} integration are provided by the direct-write nanofabrication technology. Furthermore, the ability to control the thickness of individual FIBID/FEBID layers with an accuracy better than $1$\,nm \cite{Por17nan} should allow for the fabrication of supercondcutor/insulator superlattices for studying quantum interference, commensurability effects \cite{Dob18nac} as well as photonic crystals with superconducting layers \cite{Lyu09joa}.

To summarize, we have experimentally demonstrated ultra-fast vortex dynamics at velocities up to $15$\,km/s in a uniform superconducting microstrip fabricated by focused ion beam induced deposition. A stable flux flow at such high velocities is a consequence of the combined effects of a strong edge barrier against a background of weak volume pinning, close-to-depairing critical currents, and fast quasiparticles relaxation in the investigated system. The distinctive feature of the direct-write Nb-C superconductor is a close-to-perfect edge barrier which orders the vortex motion at large current values and allows for the description of the spatial evolution of the FFI relying upon the edge-barrier-controlled FFI model. The spatial evolution of the FFI in this model goes essentially beyond the previously considered ``global'' instability models relying upon the Larkin-Ovchinnikov approach \cite{Lar76etp,Lar86inb,Bez92pcs,Doe95pcs,Shk17prb}: The presence of a current density gradient
due to the edge barrier for vortex entry leads to the nucleation of FFI points near the edge of the microstrip and a series of transformations in the vortex lattice occurs with increase of the vortex velocity. In all, the observed high vortex velocities in Nb-C-FIBID make accessible studies of far-from-equilibrium superconductivity and vortex matter driven by large currents, opening prospects for Cherenkov-like generation of other excitations by the fast-moving vortex lattice. In addition, the small electron diffusion coefficient $D \approx 0.5$\,cm$^{2}$s$^{-1}$, the low superconducting transition temperature $T_\mathrm{c}=5.6$\,K and high $I_\mathrm{c}$ values exceeding $70\%$ of the depairing current render Nb-C-FIBID to be an interesting candidate material for fast single-photon detectors.

\section*{Methods}

\textbf{Sample fabrication and its structural properties.} Superconducting microstrips were fabricated by FIBID in a dual-beam scanning electron microscope (FEI Nova Nanolab 600). The substrates are Si\,(100,\,p-doped)/SiO$_2$\,(200\,nm) with lithographically defined Au/Cr contacts for electrical transport measurements. FIBID was done at $30$\,kV/10\,pA, $30$\,nm pitch and $200$\,ns dwell time employing Nb(NMe$_2$)$_3$(N-$\textit{t}$-Bu) as precursor gas. The as-deposited Nb-C-FIBID microstrips have well-defined smooth edges and a rms surface roughness of less than $0.3$\,nm, as deduced from atomic force microscopy scans in the range $1\,\mu$m$\times 1\,\mu$m. Right after the deposition, without breaking the vacuum, the microstrips were covered with an insulating Nb-C-FEBID layer with a nominal thickness of $10$\,nm, see Fig.\,\ref{f1} for the geometry. While Nb-C-FEBID structures are amorphous, Nb-C-FIBID deposits have an fcc NbC polycrystalline structure, with grains about $15$\,nm in diameter \cite{Por19acs}. The typical elemental composition in the Nb-C-FIBID microstrips is 43\,\%\,at.\,C, $29$\,\%\,at.\,Nb, 15\,\%\,at.\,Ga, and 13\,\%\,at.\,N, as inferred from energy-dispersive X-ray spectroscopy on thicker replica of the fabricated structures. For details on the microstructural characterization of Nb-C-FEBID and Nb-C-FIBID we refer to previous work \cite{Por19acs}. Experiments were done on a series of four samples. In the manuscript we report typical data for one microstrip.
\begin{figure*}[t!]
    \centering
    \includegraphics[width=0.65\linewidth]{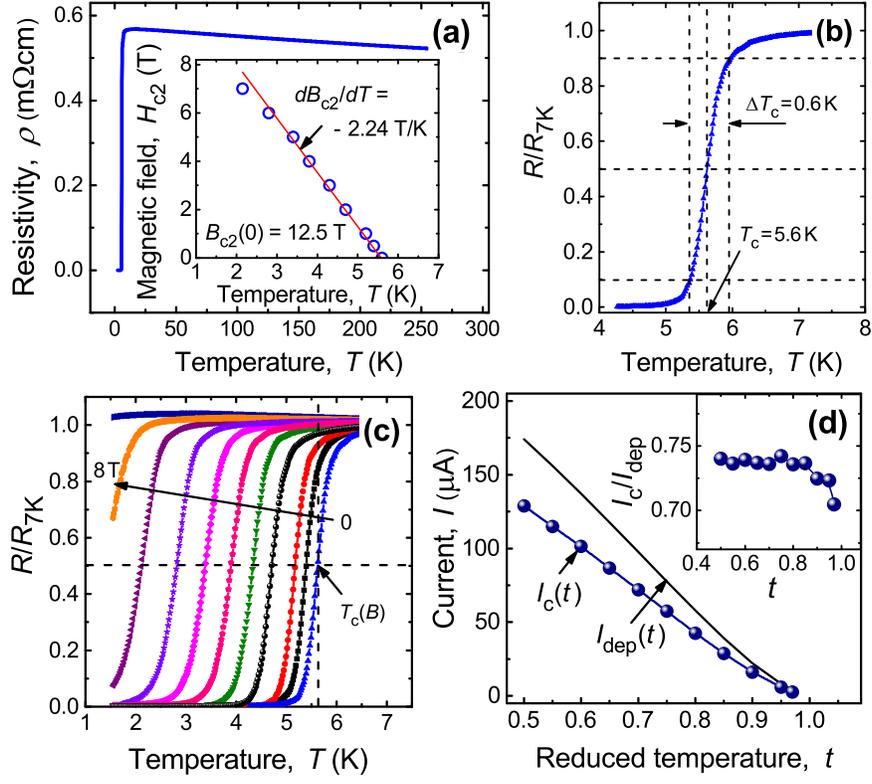}
    \caption{\textbf{Superconducting properties of the Nb-C-FIBID microstrip.} (a) Temperature dependence of the resistivity of the microstrip. Inset: Temperature dependence of the upper critical field fitted to the expression $B_\mathrm{c2}(T) = B_\mathrm{c2}(0) - (dB_\mathrm{c2}/dT)T$ with $B_\mathrm{c2}(0)=12.5$\,T and $dB_\mathrm{c2}/dT = -2.24$\,T/K. (b) Transition to the superconducting state in zero magnetic field. (c) Evolution of the superconducting transition in the presence of a magnetic field. (d) Temperature dependence of the experimentally measured critical current $I_\mathrm{c}(t)$ in comparison with the theoretically calculated depairing current $I_\mathrm{dep}(t)$. Inset: Ratio $I_\mathrm{c}/I_\mathrm{dep}$ versus reduced temperature $t$.}
        \vspace{3mm}
        \label{f2}
\end{figure*}

\textbf{Superconducting properties of the Nb-C-FIBID microstrip.} The resistive properties of the microstrip are summarized in Fig.\,\ref{f2}. The resistivity temperature dependence $\rho(T)$ is shown in Fig.\,\ref{f2}(a), where the $\rho(T)$ curve exhibits a transition from weak localization to superconductivity at $T_\mathrm{c} = 5.6$\,K. Here, the transition temperature $T_\mathrm{c}$ is determined using the $50\,\%$ resistance drop criterion, as illustrated in Fig.\,\ref{f2}(b). The resistivity at $7\,$K is $\rho_\mathrm{7K} = 572\,\mu\Omega$cm and the width of the superconducting transition, defined as the temperature difference between the $10\%$ and $90\%$ resistivity values at the transition, amounts to $\Delta T_\mathrm{c} \approx 0.6$\,K. Application of a magnetic field $B$ leads to a decrease of $T_\mathrm{c}$ and a transition broadening, and we use the same $50\,\%$ resistance drop criterion to deduce the temperature dependence of the upper critical field $B_\mathrm{c2}(T)$ shown in Fig.\,\ref{f2}(c). Near $T_\mathrm{c}$, the critical field slope $dB_{\mathrm{c}2}/dT|_{T_\mathrm{c}} = -2.24$\,T/K corresponds, in the dirty superconductor, to the electron diffusivity $D = -4 k_\mathrm{B}/[\pi e (dB_{\mathrm{c}2}/dT|_{T_\mathrm{c}})] \approx 0.49$\,cm$^2$/s. The coherence length and the penetration depth at zero temperature are estimated \cite{Kor18pra,Kes83prb} as $\xi(0) = \sqrt{\hbar D /\Delta(0)} = 6.5$\,nm  and $\lambda(0) = 1.05\cdot10^{-3} \sqrt{\rho_\mathrm{7K} k_\mathrm{B}/\Delta(0)} \approx 800\,$nm. By employing the $100\,\mu$V voltage drop criterion, from the current-voltage ($I$-$V$) curves we deduce the critical currents at zero field $I_\mathrm{c}(0.75T_\mathrm{c}) = 58\,\mu$A and $I_\mathrm{c}(0.9T_\mathrm{c}) = 16\,\mu$A. We assume that the temperature dependence of the depairing current can be described by the expression $I_\mathrm{dep}(T) = I_\mathrm{dep}(0) (1 - (T/T_\mathrm{c})^{2})^{3/2}$ with the prefactor $I_\mathrm{dep}(0) = 0.74 w [\Delta(0)]^{3/2}/(e R_\square \hbar D)$, which is justified for dirty superconductors \cite{Rom82prb,Cle12prb,Kor18pra}. Here, $\Delta(0)$ is the superconducting energy gap at zero temperature, $e$ the electron charge, and $R_\square$ the sheet resistance. With the assumed BCS ratio $\Delta(0) \approx 1.76 k_\mathrm{B}T_\mathrm{c}$ we obtain $I_\mathrm{dep}(0)\approx 268\,\mu$A. The calculated dependence $I_\mathrm{dep}(T)$ is compared with the experimentally measured $I_\mathrm{c}(T)$ in Fig.\,\ref{f2}(d). We note that $I_\mathrm{c}$ varies between $0.7I_\mathrm{dep}\lesssim I_\mathrm{c} \lesssim 0.74I_\mathrm{dep}$ in the temperature range $0.5<t<1$, where $\tau = T/T_\mathrm{c}$ is the reduced temperature.

\subsection{Time-dependent Ginzburg-Landau simulations.}
To study the evolution of the superconducting order parameter we numerically solve the modified TDGL equation \cite{Vod17pra}
\begin{eqnarray*}
    \frac{\pi\hbar}{8k_\mathrm{B}T_\mathrm{c}} \left(\frac{\partial }{\partial t}+\frac{2ie\varphi}{\hbar} \right) \Delta= \nonumber
    \\
    =\xi^2_\mathrm{mod}\left( \nabla -i\frac{2e}{\hbar c}A\right)^2\Delta+\left(1-\frac{T_\mathrm{e}}{T_\mathrm{c}}-\frac{|\Delta|^2}{\Delta_{mod}^2}\right)\Delta+
    \nonumber
    \\
    +i\frac{(\mathrm{div}\mathbf{j}_\mathrm{s}^{Us}-\mathrm{div}\mathbf{j}_\mathrm{s}^{GL})}{|\Delta|^2}\frac{e\Delta\hbar D}{\sigma_\mathrm{n}\sqrt{2}\sqrt{1+T_\mathrm{e}/T_\mathrm{c}}},
\end{eqnarray*}
where $\xi^2_\mathrm{mod}=\pi\sqrt{2}\hbar D/(8k_\mathrm{B}T_\mathrm{c}\sqrt{1+T_\mathrm{e}/T_\mathrm{c}})$,
$\Delta_\mathrm{mod}^2=(\Delta_0\tanh(1.74\sqrt{T_\mathrm{c}/T_\mathrm{e}-1}))^2/(1-T_\mathrm{e}/T_\mathrm{c})$,
$A$ is the vector potential,
$\varphi$ is the electrostatic potential,
$D$ is the diffusion coefficient,
$\mathbf{j}_\mathrm{s}^{Us}$ and
$\mathbf{j}_s^{GL}$ are the superconducting current densities in the Usadel and Ginzburg-Landau models (see Eqs. (33,\,34) in Ref. \cite{Vod17pra}), and $\sigma_\mathrm{n}=2e^2DN(0)$ is the normal-state conductivity with $N(0)$ being the single-spin density of states at the Fermi level.

The electron and phonon temperatures, $T_\mathrm{e}$ and $T_\mathrm{p}$, respectively, are found from the solution of following equations
\begin{eqnarray*}
    \frac{\partial}{\partial t}\left(\frac{\pi^2k_\mathrm{B}^2N(0)T_\mathrm{e}^2}{3}-
    \mathcal{E}_0\mathcal{E}_\mathrm{s}(T_\mathrm{e},|\Delta|)\right) = \nonumber
        \\
    = \nabla k_\mathrm{s} \nabla T_\mathrm{e}-\frac{96\zeta(5)N(0)k_\mathrm{B}^2}{\tau_0}\frac{T_\mathrm{e}^5-
    T_\mathrm{p}^5}{T_\mathrm{c}^3}+ j E,
        \\
    \frac{\partial T_\mathrm{p}^4}{\partial t}=-\frac{T_\mathrm{p}^4- T^4}{\tau_\mathrm{esc}}+\gamma\frac{24\zeta(5)}{\tau_0}\frac{15}{\pi^4}\frac{T_\mathrm{e}^5-T_\mathrm{p}^5}{T_\mathrm{c}},
\end{eqnarray*}
where $\mathcal{E}_0=4N(0)(k_\mathrm{B}T_\mathrm{c})^2$, $\mathcal{E}_0\mathcal{E}_\mathrm{s}(T_\mathrm{e},|\Delta|)$ is the change in the
energy of electrons due to transition to the superconducting state (see Eq. (26) in \cite{Vod17pra}), $k_\mathrm{s}$ is the heat conductivity in the superconducting state
\begin{equation*}
    k_\mathrm{s}=k_\mathrm{n}\left(1-\frac{6}{\pi^2(k_\mathrm{B}T_\mathrm{e})^3}\int_0^{|\Delta|}\frac{\epsilon^2
    e^{\epsilon/k_\mathrm{B}T_\mathrm{e}}d\epsilon}{(e^{\epsilon/k_\mathrm{B}T_\mathrm{e}}+1)^2}\right),
\end{equation*}
$k_\mathrm{n} = 2D\pi^2k_\mathrm{B}^2N(0)T_\mathrm{e}/3$ is the heat conductivity in the normal state, the term $jE$ describes Joule dissipation, and $\tau_\mathrm{esc}$ is the escape time of nonequilibrium phonons to the substrate. The parameter $\gamma$ is defined as
$\gamma= \displaystyle\frac{8\pi^2}{5}\displaystyle\frac{C_\mathrm{e}(T_\mathrm{c})}{C_\mathrm{p}(T_\mathrm{c})}$, where $C_\mathrm{e}(T_\mathrm{c})$ and $C_\mathrm{p}(T_\mathrm{c})$ are the heat capacities of electrons and phonons at $T=T_\mathrm{c}$, and the characteristic time $\tau_0$ controls the strength of the electron-phonon and phonon-electron scattering (see Eqs. (3),\,(4),\,(6), and\,(7) in Ref. \cite{Vod17pra}).

To find the electrostatic potential $\varphi$, we also solve the current continuity equation
\begin{equation*}
    \mathrm{div}(\mathbf{j}_\mathrm{s}^{Us}+\mathbf{j}_\mathrm{n})=0,
\end{equation*}
where $\mathbf{j}_\mathrm{n}=-\sigma_\mathrm{n}\nabla \varphi$ is the normal current density.

Values of the parameters $\gamma=9$ and $\tau_0=925$\,ns used in the calculations are estimates for NbN. Their variation only leads to quantitative changes in the $I$-$V$ curves.

At the edges where vortices enter and exit the microstrip we use the boundary conditions $\mathbf{j}_\mathrm{n}|_\mathrm{n}=\mathbf{j}_\mathrm{s}|_\mathrm{n}=0$ and $\partial T_\mathrm{e}/\partial\mathrm{n}=0$, $\partial |\Delta|/\partial n=0$ while at the edges along the current direction $T_\mathrm{e}=T$, $|\Delta|=0$, $\mathbf{j}_\mathrm{s}|_\mathrm{n}=0$, $\mathbf{j}_\mathrm{n}|_\mathrm{n}=I/wd$. The latter boundary conditions model the contact of the superconducting strip with a normal reservoir being in equilibrium. This choice provides a way ``to inject'' the current into the superconducting microstrip in the modeling. The modeled length of the microstrip is $L=4w$.

In the considered model, the penetration length of the electric field $L_\mathrm{E}$ is about the coherence length $\xi(T)$, which is a consequence of $\tau_\mathrm{ee} \ll \tau_\mathrm{ep}$. If $\tau_\mathrm{ee}\gtrsim \tau_\mathrm{ep}$, then $L_\mathrm{E}$ can be considerably larger than
$\xi(T)$. In general, $L_\mathrm{E}$ stipulates the stability of the phase slip process in 1D superconducting wires at larger currents \cite{Vod05prba}. In the case of vortex rivers (phase slip lines with vortices) it should also lead to their stability at larger currents, providing a critical velocity of Abrikosov vortices close to the velocity of Josephson vortices, which could explain the experimentally observed $v^{\ast} \gtrsim 10$\,km/s. Within the framework of the considered model, a larger $L_\mathrm{E}$ can be modeled by a smaller numerical coefficient at the time derivative
$\partial \Delta/\partial t$. This simultaneously leads to a decrease of the relaxation time of $|\Delta|$, which also leads to an increase of $v^\ast$. For instance, a fivefold decrease of this coefficient (that corresponds to an increase of $L_\mathrm{E}$ by a factor of $\approx \sqrt{5}$) results in a twofold increase of $V^\ast$ and $v^\ast$ and a small decrease of $I^\ast$ at $B=0.1 B_0$. One can also see that in this case vortex rivers are well formed at $I=I^\ast$ and Abrikosov vortices are closer to Abrikosov-Josephson vortices because of the stronger suppression of the order parameter along the vortex river, leading to higher instability velocities.

\section*{Appendix}
\textbf{Comparison of the instability parameters with the Bezuglyj-Shklovskij model}.
In their work \cite{Bez92pcs}, Bezuglyj and Shklovskij introduced a scaling law for the electric field strength $E^\ast$ and the current density $j^\ast$ at the instability point
\begin{equation}
\label{eLO}
    \displaystyle\frac{E^\ast}{E_0}= (1 - t(b))\left(\frac{j^\ast}{j_0}\right)^{-1}.
\end{equation}
In Eq. (\ref{eLO}), parameters $E_0$ and $j_0$ are defined as
\begin{equation}
\label{eLOparam}
    \begin{array}{lll}
    \displaystyle E_0 = 1.02 B_\mathrm{T}(D/\tau_\epsilon)^{1/2}(1-T/T_\mathrm{c})^{1/4},
    \\[2mm]
    \displaystyle j_0 = 2.62 (\sigma_\mathrm{n}/e)(D\tau_\epsilon)^{-1/2}k_\mathrm{B} T_\mathrm{c}(1-T/T_\mathrm{c})^{3/4},
    \end{array}
\end{equation}
$t =[1+b+(b^2 +8b+4)^{1/2}]/3(1+2b)$ and $b = B / B_\mathrm{T}$ is the magnetic field normalized by the parameter
\begin{equation}
\label{eBt}
    B_\mathrm{T} = 0.374 k_\mathrm{B}^{-1}eR_{\square}h\tau_\epsilon.
\end{equation}
In Eq. (\ref{eBt}), $h$ is the heat removal coefficient and $\tau_\epsilon$ is the quasiparticle energy relaxation time. The parameter $B_\mathrm{T}$ separates the region of small fields $B\lesssim B_\mathrm{T}$ at which heat removal is fast enough and the instability is of non-thermal nature from the region of large fields $B_\mathrm{T} \lesssim B \lesssim 0.4 B_\mathrm{c2}$ with insufficient heat removal and the heating mechanism dominating the instability.
\begin{figure}[t!]
    \centering
    \includegraphics[width=0.8\linewidth]{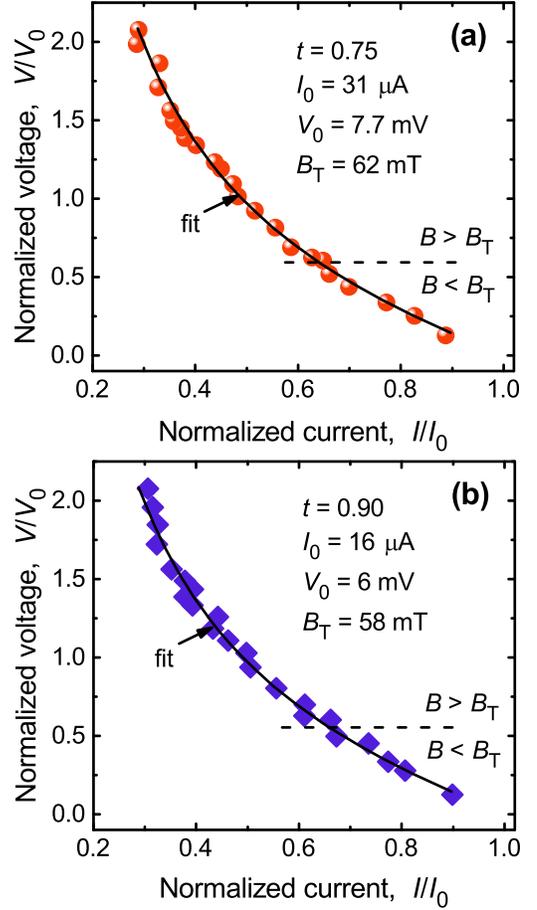}
    \caption{Complete set of instability points at $0.75T_\mathrm{c}$ (a) and $0.9T_\mathrm{c}$ (b). Symbols: experiment; solid lines: fits to Eq. (\ref{eLO}) with the fitting parameters as indicated.}
    \label{fBS}
\end{figure}

The curves calculated by Eqs. (\ref{eLO}) and (\ref{eLOparam}) are shown in Fig.\,\ref{fBS}(a) and (b) by solid lines. The theoretical curves nicely fit the experimentally measured instability points in the normalized voltage $V^\ast/V_0 = E^\ast/E_0$ versus normalized current $I^\ast/I_0 = j^\ast/j_0$ representation with the fitting parameters $B_\mathrm{T} = 62$\,mT, $V_0 = 7.7$\,mV and $I_0=31\,\mu$A at $0.75T_\mathrm{c}$, and $B_\mathrm{T} = 58$\,mT, $V_0 = 6.0$\,mV and $I_0=16\,\mu$A at $0.9T_\mathrm{c}$. Here, the field- and temperature-dependent instability currents $I^\ast$ and voltages $V^\ast$ are determined from the $I$-$V$ curves. From the specific power at the instability point, $P_0 = j_0 E_0 =(h/d)(T_\mathrm{c} - T)$ \cite{Bez92pcs,Lef99pcs,Per05prb,Bez19prb}, following from Eqs. (\ref{eLO})--(\ref{eBt}) with $\sigma_n = 1/(R_\square d)$, one can deduce the heat removal coefficient $h\approx 2.6$\,WK$^{-1}$cm$^{-2}$. Substitution of $h$ and $B_\mathrm{T}$ into Eq. (\ref{eBt}) yields the energy relaxation time $\tau_{\epsilon} \approx 1.4$\,ps. if one associates $\tau_{\epsilon}$ with the electron-phonon scattering time $\tau_\mathrm{ep}$ in the Larkin-Ovchinnikov model, the deduced $\tau_{\epsilon}$ is at least one order of magnitude smaller than one could expect from $\tau_{\epsilon}$ found in similar low-$T_\mathrm{c}$ highly disordered superconductors \cite{Bab04prb,Sid18prb,Sid19arx}.

The deduced heat removal coefficient $h\approx 2.6$\,WK$^{-1}$cm$^{-2}$ is of the same order of magnitude as for dirty Nb films on sapphire substrates \cite{Per05prb}, and it is one to two orders of magnitude smaller than $h$ values for epitaxial BSSCO films on SrTiO$_3$ substrates \cite{Xia98prb1} and epitaxial YBaCuO films on sapphire substrates \cite{Xia98prb2}. We assume that the presence of the Nb-C-FEBID layer on top of Nb-C-FIBID may have improved the effective heat removal from the sample.

OD acknowledges the German Research Foundation (DFG) for support through Grant No 374052683 (DO1511/3-1).
            DV acknowledges the Russian Science Foundation for support through grant No 17-72-30036.
            AVC acknowledges support within the ERC Starting Grant No 678309 MagnonCircuits.
            Further, this work was supported by the European Cooperation in Science and Technology via COST Action CA16218 (NANOCOHYBRI).
            Support through the Frankfurt Center of Electron Microscopy (FCEM) is gratefully acknowledged.


%

\end{document}